\documentclass[runningheads]{llncs}
\usepackage{etex}
\usepackage{tikz}
\usetikzlibrary{calc} 						
\usetikzlibrary{shapes,arrows}				
\usetikzlibrary{shapes.multipart}			
\usetikzlibrary{positioning}
\usepackage[all,cmtip]{xy}					
\usepackage{tikz-qtree,tikz-qtree-compat}

\usepackage{url}
\usepackage{turnstile}
\usepackage{epsfig}
\usepackage{amssymb}
\usepackage[all]{xypic}
\usepackage{amsmath}
\usepackage{latexsym}
\usepackage{cancel}

\long\def\comment#1{}

\newcommand{\formulasOn}[1]{\mathit{Form}_{#1}}
\newcommand{\SPformulas}{\formulasOn{\SSymbols,\SPredicates}}

\newcommand{\SAlgebra}{\cA}
\newcommand{\SStructure}{\cA}

\newcommand{\SModel}{\cA}
\newcommand{\InitialModel}{\cI}

\newcommand{\SemDomain}{\cA}

\newcommand{\DomainOf}[1]{\mathit{dom}(#1)}
\newcommand{\ModelDomain}{\cA}

\newcommand{\IF}{\Leftarrow}

\newcommand{\subtermeq}{\unrhd}

\newcommand{\AGES}{{\sf AGES}}

\newcommand{\Maude}{{\sf Maude}}

\newcommand{\bigfrac}[2]{
\begin{array}[b]{c}
\displaystyle #1\\\hline\displaystyle #2
\end{array}}

\newcommand{\bigfracn}[3]{
\begin{array}[b]{c}
\displaystyle #1 \\\hline\displaystyle #2 
\end{array}
\hbox to 0pt{\raisebox{0.7em}{{\tiny (#3)}}}
}

\makeatletter
\newenvironment{prog}{\vspace{0.7ex}\par
\setlength{\parindent}{0.7cm}
\obeylines\@vobeyspaces\tt}{\vspace{0.0ex}\noindent
}
\makeatother
\newcommand{\startprog}{\begin{prog}}
\newcommand{\stopprog}{\end{prog}\noindent}

\makeatletter
\newenvironment{smallprog}{\vspace{0.7ex}\par
\setlength{\parindent}{0.7cm}
\obeylines\@vobeyspaces\tt\small}{\vspace{0.7ex}\noindent
}
\makeatother
\newcommand{\fstartprog}{\begin{smallprog}}
\newcommand{\fstopprog}{\end{smallprog}\noindent}

\makeatletter
\newenvironment{nismallprog}{\vspace{0.7ex}\par
\setlength{\parindent}{0.0cm}
\obeylines\@vobeyspaces\tt\small}{\vspace{0.7ex}\noindent
}
\makeatother
\newcommand{\fnistartprog}{\begin{nismallprog}}
\newcommand{\fnistopprog}{\end{nismallprog}\noindent}

\newcommand{\pr}[1]{\mbox{\tt #1}}   

\newcommand{\ol}[1]{\overline{#1}}  

\def\defemb#1#2{\expandafter\def\csname #1\endcsname
{\relax\ifmmode #2\else\hbox{$#2$}\fi}}
\defemb{cA}{{\cal A}}
\defemb{cB}{{\cal B}}
\defemb{cC}{{\cal C}}
\defemb{cD}{{\cal D}}
\defemb{cE}{{\cal E}}
\defemb{cF}{{\cal F}}
\defemb{cG}{{\cal G}}
\defemb{cH}{{\cal H}}
\defemb{cI}{{\cal I}}
\defemb{cJ}{{\cal J}}
\defemb{cL}{{\cal L}}
\defemb{cM}{{\cal M}}
\defemb{cN}{{\cal N}}
\defemb{cO}{{\cal O}}
\defemb{cP}{{\cal P}}
\defemb{cQ}{{\cal Q}}
\defemb{cR}{{\cal R}}
\defemb{cS}{{\cal S}}
\defemb{cT}{{\cal T}}
\defemb{cU}{{\cal U}}
\defemb{cW}{{\cal W}}
\defemb{cV}{{\cal V}}
\defemb{cX}{{\cal X}}
\defemb{cZ}{{\cal Z}}

\newcommand{\SSignature}{{\Symbols}}
\newcommand{\SPSignature}{{\Omega}}

\newcommand{\SSymbols}{{\Symbols}}
\newcommand{\SPredicates}{{\Pi}}

\newcommand{\STerms}{{\TermsOn{\SSymbols}{\Variables}}}
\newcommand{\GSTermsOn}[1]{{\GTermsOn{#1}}}
\newcommand{\GSTerms}{{\GSTermsOn{\SSymbols}}}

\newcommand{\Symbols}{{\cF}}

\newcommand{\Variables}{{\cX}}
\newcommand{\TermsOn}[2]{{\cT(#1,#2)}}
\newcommand{\GTermsOn}[1]{{\cT(#1)}}

\newcommand{\toppos}{{\Lambda}} 

\newcommand{\naturals}{\mathbb{N}}

\newcommand{\toStarPosWith}[2]{{\;\mbox{$\stackrel{#1}{\longrightarrow}\hspace{.1cm}\hspace{-.2cm}^*_{#2}\,$}}}

\newcommand{\exSymbType}[1]{\mathsf{#1}}

\newcommand{\Fa}{\exSymbType{a}}

\newcommand{\Fb}{\exSymbType{b}}
\newcommand{\Fc}{\exSymbType{c}}

\newcommand{\Fd}{\exSymbType{d}}

\newcommand{\Ff}{\exSymbType{f}}

\newcommand{\Fg}{\exSymbType{g}}

\newcommand{\Fh}{\exSymbType{h}}

\newcommand{\Fs}{\exSymbType{s}}

\newcommand{\Ftrue}{\exSymbType{true}}
\newcommand{\Fz}{\exSymbType{0}}

\newcommand{\FG}{\exSymbType{G}}

\newcommand{\FP}{\exSymbType{P}}

\tikzstyle{decision} = [diamond, draw, fill=yellow!20, text width=5em, text badly centered, minimum height=4em, inner sep=0pt, aspect=2]
\tikzstyle{block} = [rectangle, draw,fill=blue!20, text width=5em, text centered, minimum height=4em, rounded corners]
\tikzstyle{cloud} = [ellipse, draw,fill=red!20, text width=5em, text centered, minimum height=4em]
\tikzstyle{line} = [draw, -latex']
\tikzstyle{blockR} = [rectangle, draw, fill=red!20, text centered, minimum height=4em, rounded corners, minimum height=0.75cm]
\tikzstyle{blockB} = [rectangle, draw, fill=blue!20, text centered, minimum height=4em, rounded corners, minimum height=0.75cm]
\tikzstyle{blockG} = [rectangle, draw, fill=green!20, text centered, minimum height=4em, rounded corners, minimum height=0.75cm]
\tikzstyle{blockY} = [rectangle, draw, fill=yellow!20, text centered, minimum height=4em, rounded corners, minimum height=0.75cm]
\tikzstyle{blockW} = [rectangle, draw, fill=white!20, text centered, minimum height=4em, rounded corners, minimum height=0.75cm]
\tikzstyle{blockK} = [rectangle, draw, fill=gray!20, text centered, minimum height=4em, rounded corners, minimum height=0.75cm]
\tikzstyle{mblockB} = [rectangle, draw, fill=blue!20, text centered, minimum height=4em, double,rounded corners, minimum height=0.75cm]
\tikzstyle{mblockW} = [rectangle, draw, fill=white!20, text centered, minimum height=4em, double,rounded corners, minimum height=0.75cm]
\tikzstyle{mblockR} = [rectangle, draw, fill=red!20, text centered, minimum height=4em, double,rounded corners, minimum height=0.75cm]
\tikzstyle{mblockG} = [rectangle, draw, fill=green!20, text centered, minimum height=4em, double,rounded corners, minimum height=0.75cm]

\tikzstyle{circleW} = [circle, draw, fill=white!20, text centered, minimum height=4em, rounded corners, minimum height=0.75cm]
\tikzstyle{triangleW} = [isosceles triangle, draw, fill=white!20, text centered, shape border rotate = -90, isosceles triangle stretches]
\tikzstyle{triangleG} = [isosceles triangle, draw, fill=green!20, text centered, shape border rotate = -90, isosceles triangle stretches]
\tikzstyle{triangleR} = [isosceles triangle, draw, fill=red!20, text centered, shape border rotate = -90, isosceles triangle stretches]

\begin{document}

\title{A Semantic Approach to the Analysis of Rewriting-Based Systems%
\thanks{Partially supported by the EU
    (FEDER), Spanish
    MINECO project TIN2015-69175-C4-1-R and
     GV project PROMETEOII/2015/013.  
    }
}
\author{
Salvador Lucas 
}

\institute{
DSIC,  Universitat Polit\`ecnica de Val\`encia, Spain\\
\url{http://users.dsic.upv.es/~slucas/}
}

\titlerunning{Analysis of Rewriting-Based Systems}

\authorrunning{Lucas}

\maketitle

\begin{abstract}
Properties expressed as the provability of a first-order sentence 
can be \emph{disproved} by just finding a model of the negation of the sentence.
This fact, however, is meaningful  in restricted cases only, depending on the shape of the sentence
and the class of systems at stake.

In this paper we show that a number of interesting properties of rewriting-based systems can be investigated 
in this way, including 
infeasibility and non-joinability of critical pairs in (conditional) rewriting, non-loopingness of conditional
rewrite systems, or the secure access to protected pages of a web site modeled as an order-sorted rewrite theory.
Interestingly, this uniform, semantic approach succeeds when specific  techniques developed
to deal with the aforementioned problems fail. 
\end{abstract}


\setcounter{page}{1}

\section{Introduction}
\label{sec:intro}
First-Order Logic is 
an appropriate language to express the semantics of computational
systems and also the (claimed) properties of such computational systems
\cite{ChaLee_SymbolicLogicAndMechanicalTheoremProving_1973}.
In this paper we explore 
the use of first-order logic in the analysis of rewriting-based systems,
including Term Rewriting Systems (TRSs, \cite{BaaNip_TermRewAllThat_1998}), 
\emph{Conditional} TRSs (CTRSs, \cite{BerKlo_ConditionalRewriteRulesConfluenceAndTermination_JCSS86,%
DerOka_RationaleForConditionalEqProgramming_TCS90,Ohlebusch_AdvTopicsTermRew_2002}),
\emph{Membership Equational Programs} \cite{DuranEtAl_ProvOpTermMEqProg_HOSC08,%
Meseguer_MembershipAlgebraAsALogicalFrameworkForEquationalSpecification_WADT97},
and more general rewriting-based formalisms 
\cite{BruMes_SemFoundGRT_TCS06,%
GogMes_ModelsAndEqualityForLogicalProgramming_TAPSOFT87,%
Meseguer_20YearsRewLogic_JLAP12}.
The insertion of a  `rewriting-based system' $\cR$ into First-Order Logic is made as (the specification of) a 
\emph{Horn theory}, i.e., a set of sentences $\ol{\cR}$ which are universally quantified implications $A_1\wedge\cdots\wedge A_n\Rightarrow B$ 
 for some $n\geq 0$ where $A_i$, $1\leq i\leq n$ and $B$ are \emph{atoms} corresponding to predicate symbols $\to$, $\to^*$, etc.
Such a \emph{Horn theory} is usually obtained from the \emph{operational semantics} of the system usually given by means of 
some \emph{inference rules}.
\begin{example}\label{ExLoopingnessInCTRSs}
Consider the following CTRS $\cR$:
\begin{eqnarray}
\Fb & \to & \Fa\label{ExLoopingnessInCTRSs_rule1}\\
\Fa & \to & \Fb \IF \Fc \to \Fb\label{ExLoopingnessInCTRSs_rule2}
\end{eqnarray}
Its associated Horn theory $\ol{\cR}$ (using predicate symbols $\to$ and $\to^*$) is:
\\[-0.6cm] 
\begin{tabular}{c@{\hspace{-0.6cm}}c}
\hspace{-1cm}
\begin{minipage}[t]{.58\linewidth}
\begin{eqnarray}
(\forall x)\: x & \to^* & x\label{ExLoopingnessInCTRSs_HornTh_reflexivity}\\
(\forall x,y,z)\: x \to y \wedge y \to^* z \Rightarrow x & \to^* & z\label{ExLoopingnessInCTRSs_HornTh_transitivity}
\end{eqnarray}\nonumber
\end{minipage} & 
\begin{minipage}[t]{.55\linewidth}
\begin{eqnarray}
\Fb & \to & \Fa\label{ExLoopingnessInCTRSs_HornTh_rule1}\\
\Fc \to^* \Fb \Rightarrow \Fa & \to & \Fb \label{ExLoopingnessInCTRSs_HornTh_rule2}
\end{eqnarray}\nonumber
\end{minipage}
\end{tabular}\\[0.3cm]
Sentence (\ref{ExLoopingnessInCTRSs_HornTh_reflexivity}) 
corresponds to \emph{reflexivity} of the many-step rewrite relation $\to^*$ and (\ref{ExLoopingnessInCTRSs_HornTh_transitivity})
is usually called \emph{transitivity}, although it actually says how the one-step rewrite relation $\to$ and the many-step 
relation are related.
Finally, (\ref{ExLoopingnessInCTRSs_HornTh_rule1}) and (\ref{ExLoopingnessInCTRSs_HornTh_rule2}) describe the CTRS at stake.
\end{example}
In this setting, our approach goes back to Floyd, Hoare, and Manna's early work on proving program properties using first-order logic:
we can use logical formulas to describe the execution of a program 
and
then other formulas describe the property of interest \cite[Chapter 10]{ChaLee_SymbolicLogicAndMechanicalTheoremProving_1973}.
However, the natural idea of using the notion of \emph{logical consequence} $\ol{\cR}\models\varphi$ 
(i.e., that $\varphi$ is satisfied in \emph{every model} of $\ol{\cR}$)
as a formal definition of \emph{``system $\cR$ has property $\varphi$''}
may fail to work.

\begin{example}\label{ExLoopingnessInCTRSs_NonRewriting}
(Continuing Example \ref{ExLoopingnessInCTRSs})
Note that $\Fa$ does \emph{not} rewrite into $\Fb$ because the conditional part of rule (\ref{ExLoopingnessInCTRSs_rule2}) cannot be satisfied: 
$\Fc$ cannot be 
rewritten into $\Fb$.
Following the aforementioned `natural approach', we are tempted to formalize this as follows:
$\ol{\cR}\models\neg(\Fa\to\Fb)$ holds, i.e.,
every model of $\ol{\cR}$ satisfies $\neg(\Fa\to\Fb)$.
However, an interpretation of the constant symbols $\Fa$ and $\Fb$ as $0$, with $\to$ and $\to^*$ interpreted as the \emph{equality} 
satisfies $(\ref{ExLoopingnessInCTRSs_HornTh_reflexivity})-(\ref{ExLoopingnessInCTRSs_HornTh_rule2})$ 
(i.e., it is a model of $\ol{\cR}$), but $\neg(\Fa\to\Fb)$
does \emph{not} hold. Thus, $\ol{\cR}\models\neg(\Fa\to\Fb)$ does \emph{not} hold!
\end{example}
This `mismatch' between the expressivity of pure first-order logic and the intended meaning of logic sentences referred to the \emph{computational
logic} describing a given computational system is usually avoided by the assumption that sentences expressing program properties
should be \emph{checked} with respect to a given \emph{canonical model} only \cite[Chapter 4]{Clark_PredicateLogicAsAComputationalFormalism_TR79}.
For instance, the problem in Example \ref{ExLoopingnessInCTRSs_NonRewriting} disappears if we assume that 
$\neg(\Fa\to\Fb)$ must hold in the \emph{least Herbrand model} $\cH_\cR$ of $\ol{\cR}$ only.
In $\cH_\cR$, $\to$ and $\to^*$ are interpreted precisely as the 
sets $(\to)^{\cH_\cR}$ and $(\to^*)^{\cH_\cR}$  of pairs $(s,t)$ of ground terms $s$ and $t$ such that $s\to_\cR t$ and $s\to^*_\cR t$, respectively.
Then, we indeed have $\cH_\cR\models\neg(\Fa\to\Fb)$, which is agreed to be the \emph{intended meaning} of the logic expression $\neg(\Fa\to\Fb)$.

In general, the (standard) least Herbrand model $\cH$ of a Horn theory is not computable\footnote{For instance, in the rewriting setting, it is well-known that rechability of terms, i.e., whether $s\to^*_\cR t$ for given terms $s$ and $t$, is undecidable 
(Post's correspondence problem is a particular case). This means that 
$\cH_\cR\models s\to^*t$ is undecidable too.}. Thus, the practical verification of properties $\varphi$ as
satisfiability in $\cH$, i.e., $\cH\models\varphi$, is not possible, in general.
In this paper we show that the class of properties $\varphi$ which can be written 
as the \emph{existential closure of a positive boolean combination of atoms} can be \emph{disproved}
(with regard to the least Herbrand model of a Horn theory $\cS$) by showing the satisfiability 
of $\neg\varphi$ in an \emph{arbitrary} model $\SStructure$ of $\cS$, i.e., by proving $\SStructure\models\neg\varphi$.
When this approach is applied to rewriting-based systems $\cR$ and the associated Horn theory $\ol{\cR}$, a
 number of interesting properties (some of them already considered in the literature) can be expressed and disproved in this way.
Some examples are given in Figure \ref{FigSomeFOpropertiesRewritingBasedSystems}, where $s$ and $t$ denote  \emph{ground terms}, $s_1,\ldots,s_n,t_1,\ldots,t_n$ denote arbitrary terms with variables in $\vec{x}$ (in the feasibility property, see \cite{LucGut_ASemanticCriterionForProvingInfeasibitlityInConditionalRewriting_IWC17}) 
and $\subtermeq$ is the
\emph{subterm relation}.

\begin{figure}
\begin{center}
\begin{tabular}{|@{~}c@{~}|@{~}c@{~}|}
\hline
\emph{Property} & $\varphi$\\
\hline
\emph{Reach}able & $s\to^*t$\\
\emph{Feas}ible & $(\exists \vec{x})s_1\to^*t_1\wedge\cdots\wedge s_n\to^*t_n$\\
\emph{Join}able &  $(\exists x)~(s\to^*x\wedge t\to^*x)$\\
\emph{Red}ucible & $(\exists x)~t\to x$\\
\emph{Conv}ertible & $s\to t \vee t\to s$\\
\emph{Cycl}ing \emph{t}erm & $(\exists x)~t\to x\wedge x\to^*t$\\
\emph{Cycl}ing system & $(\exists x,y)~x\to y\wedge y\to^*x$\\
\emph{Loop}ing \emph{t}erm & $(\exists x,y)~t\to x\wedge x\to^*y\wedge y\subtermeq t$\\
\emph{Loop}ing system & $(\exists x,y,z)~x\to y\wedge y\to^*z\wedge z\subtermeq x$\\
\hline
\end{tabular}
\end{center}
\caption{Some properties about rewriting-based systems}
\label{FigSomeFOpropertiesRewritingBasedSystems}
\end{figure}
\begin{example}\label{ExLoopingnessInCTRSs_ProofOfIrreducible}
(Continuing Example \ref{ExLoopingnessInCTRSs_NonRewriting}) 
The fact that $\Fa$ rewrites into $\Fb$ (i.e., $\Fa\to_\cR\Fb$) can be \emph{disproved} if there is 
a model $\SStructure$ 
of (\ref{ExLoopingnessInCTRSs_HornTh_reflexivity})-(\ref{ExLoopingnessInCTRSs_HornTh_rule2}) 
satisfying $\neg(\Fa\to\Fb)$.
The interpretation $\SStructure$ with domain $\naturals$, 
interpreting both $\Fa$ and $\Fc$ as $1$, $\Fb$ as $2$, $\to$ as $>_\naturals$ and $\to^*$ as $\geq_\naturals$ is a model of
$\{(\ref{ExLoopingnessInCTRSs_HornTh_reflexivity})-(\ref{ExLoopingnessInCTRSs_HornTh_rule2})\}\cup\{\neg(\Fa\to\Fb)\}$.
This proves that $\Fa\not\to_\cR\Fb$.
\end{example}
After some preliminaries, Section \ref{SecExistentiallyClosedBooleanCombinationsOfAtoms}
presents the main result of the paper which is formulated in a standard first-order logic framework
\cite{Mendelson_IntroductionToMathematicalLogicFourtEd_1997}.
Section \ref{SecApplicatonToRewriting} explains its use in a rewriting setting.
By lack of space we mainly focus on CTRSs but other computational systems could be treated in this way.
Section \ref{SecRelatedWork} discusses some related work. Section \ref{SecConclusions} concludes.

\section{Preliminaries}

A \emph{signature with predicates}\footnote{We follow the terminology and notation in \cite{GogMes_ModelsAndEqualityForLogicalProgramming_TAPSOFT87}.} $\SPSignature$ is a pair $\SPSignature=(\SSignature,\SPredicates)$, where $\SSignature$ is a 
set of function symbols $\SSignature=\{f,g,\ldots\}$  
and $\SPredicates$ is a set of \emph{predicate symbols} $\SPredicates=\{P,Q,\ldots\}$ with $\SSymbols\cap\SPredicates=\emptyset$.
An \emph{arity} mapping $ar:\SSignature\cup\SPredicates\to\naturals$ fixes the number of arguments
for each symbol.
First-order terms $t$ and formulas $\varphi$ are built from these symbols (and an infinite set $\Variables$
of variable symbols $\Variables=\{x,y,z,\ldots\}$, which is disjoint from $\SSignature\cup\SPredicates$) in the 
usual way. 
Equations $s=t$ for terms $s$ and $t$ can also be used as \emph{atoms} if necessary, 
even without any equality symbol in $\SPredicates$.
The set of terms is denoted as $\STerms$ (the set of \emph{ground} terms, i.e., terms \emph{without variables}, is denoted as $\GSTerms$.
The set of (first-order) formulas is denoted as $\SPformulas$.

An $\SPSignature$-structure $\SStructure$ for a signature with predicates $\SPSignature$ is an interpretation 
of the function and predicate symbols in $\SPSignature$ 
as mappings $f^\SStructure,g^\SStructure,\ldots$ and relations $P^\SStructure,Q^\SStructure,\ldots$ on a given set (carrier) 
$\DomainOf{\SStructure}$, often denoted $\SemDomain$ as well.
The equality symbol has a \emph{fixed}
interpretation as the identity relation $\{(a,a)\mid a\in\SemDomain\}$ on $\SemDomain$.
 An $\SPSignature$-homomorphism between $\SPSignature$-structures $\SAlgebra$ and $\SAlgebra'$
is a mapping $h:\DomainOf{\SemDomain}\to\DomainOf{\SemDomain'}$ such that
(i) for each $k$-ary symbols $f\in\SSymbols$, and $a_1,\ldots,a_k\in\DomainOf{\SemDomain}$,
$h(f^\SAlgebra(a_1,\ldots,a_k))=f^{\SAlgebra'}(h(a_1),\ldots,h(a_k))$ and
(ii) for each $n$-ary predicate symbols $P\in\SPredicates$ and $a_1,\ldots,a_n\in\DomainOf{\SemDomain}$,
if $(a_1,\ldots,a_n)\in P^{\SStructure}$, then $(h(a_1),\ldots,h(a_n))\in P^{\SStructure'}$ \cite[Section 1.2]{Hodges_AShorterModelTheory_1997}. 
Given a  \emph{valuation mapping} $\alpha:\Variables\to\SemDomain$, the evaluation mapping
$[\_]^\alpha_\SStructure:\STerms\to \SemDomain$ 
is given by $[t]^\alpha_\SStructure=\alpha(t)$ if $t\in\Variables$ and
$[t]^\alpha_\SStructure=f^\SStructure([t_1]^\alpha_\SStructure,\ldots,[t_k]^\alpha_\SStructure)$
if $t=f(t_1,\ldots,t_k)$ (if $k=0$, then $t$ is just a constant symbol $f$).
Finally, $[\_]^\alpha_\SStructure:\SPformulas\to\mathit{Bool}$ is given by: 
\begin{enumerate}
\item $[P(t_1,\ldots,t_n)]^\alpha_\SStructure=\mathit{true}$ (with $P\in\SPredicates$) 
if and only if $([t_1]^\alpha_\SStructure,\ldots,[t_n]^\alpha_\SStructure)\in P^\ModelDomain$;
\item $[\neg\phi]^\alpha_\SStructure=\mathit{true}$ if and only if $[\phi]^\alpha_\SStructure=\mathit{false}$;
\item $[\phi\wedge\psi]^\alpha_\SStructure=\mathit{true}$ if and only if $[\phi]^\alpha_\SStructure=\mathit{true}$ and $[\psi]^\alpha_\SStructure=\mathit{true}$; 
\item $[\phi\vee\psi]^\alpha_\SStructure=\mathit{true}$ if and only if $[\phi]^\alpha_\SStructure=\mathit{true}$ or $[\psi]^\alpha_\SStructure=\mathit{true}$; 
\item $[(\forall x)\:\phi]^\alpha_\SStructure=\mathit{true}$ if and only if for all $a\in\SemDomain$, $[\phi]^{\alpha[x\mapsto a]}_\SStructure=\mathit{true}$; and
\item $[(\exists x)\:\phi]^\alpha_\SStructure=\mathit{true}$ if and only if there is $a\in\SemDomain$, such that $[\phi]^{\alpha[x\mapsto a]}_\SStructure=\mathit{true}$.
\end{enumerate}
A valuation $\alpha\in\Variables\to\SemDomain$ \emph{satisfies} a formula $\varphi$ 
in  $\SStructure$ 
(written $\SStructure\models\varphi\:[\alpha]$) if $[\varphi]^\alpha_\SStructure=\mathit{true}$.
A model for a \emph{theory} $\cS$, i.e., a set of \emph{sentences} 
(which are formulas whose variables are all \emph{quantified}), is just a structure that makes them all true, written $\SStructure\models\cS$, see \cite{Hodges_AShorterModelTheory_1997}.
Let $\mathit{Mod}(\cS)$ be the class of structures $\SStructure$ which are models of $\cS$.
A sentence $\varphi$ is a logical consequence of a theory $\cS$ (written $\cS\models\varphi$) if for all $\SStructure\in\mathit{Mod}(\cS)$,
$\SStructure\models\varphi$.
If $\varphi$ can be \emph{proved} from $\cS$ by using an appropriate calculus (e.g., the axiomatic calculus by Hilbert
\cite[Section 2.3]{Mendelson_IntroductionToMathematicalLogicFourtEd_1997}, or Gentzen's natural
deduction, see  \cite{Prawitz_NaturalDeductionAProofTheoreticalStudy_2006}), we write $\cS\vdash\varphi$.

\section{Existentially Closed Boolean Combinations of Atoms}\label{SecExistentiallyClosedBooleanCombinationsOfAtoms}

Every set $\cS$ of ground atoms has an \emph{initial model}.
 
 \begin{theorem}{\rm  \cite[Theorem 1.5.2]{Hodges_AShorterModelTheory_1997}}
 \label{TheoExistenceOfInitialModel}
 Let $\SPSignature$ be a first-order signature and $\cS$ be a set of ground atoms.
 Then, there is a structure $\InitialModel_\cS$ such that
 \begin{enumerate}
 \item $\InitialModel_\cS\models \cS$,
 \item every element of $\DomainOf{\InitialModel_\cS}$ is of the form $t^{\InitialModel_\cS}$ for some ground term $t$,
 \item if $\SStructure$ is an $\SPSignature$-structure and $\SStructure\models\cS$, then there is a unique homomorphism $h:\InitialModel_\cS\to\SStructure$.
 \end{enumerate}
 \end{theorem}
Actually, the \emph{initial structure} $\InitialModel_\cS$ (or just $\InitialModel$, if $\cS$ is understood from the context) 
which is mentioned in Theorem \ref{TheoExistenceOfInitialModel}, and also in some of the results below,
consists of the usual \emph{Herbrand Domain} of ground terms
modulo the equivalence $\sim$ generated by the equations in $\cS$ \cite[Lemma 1.5.1]{Hodges_AShorterModelTheory_1997}: 
For each ground term $t\in\GSTerms$, let $t^\sim$ be the equivalence class of $t$ under $\sim$. Then,
\begin{enumerate}
\item For each constant $c\in\SSignature$, we let $c^\InitialModel=c^\sim$.
\item For each function symbol $f\in\SSignature$ of arity $k$, define $f^\InitialModel$ by $f^\InitialModel(t_1^\sim,\ldots,t_k^\sim)=f(t_1,\ldots,t_k)^\sim$.
\item For each predicate symbol $P\in\SPSignature$ of arity $n$, define $P^\InitialModel$ as the set $\{(t_1^\sim,\ldots,t_n^\sim)\mid P(t_1,\ldots,t_n)\in\cS\}$.
\end{enumerate}
If $\cS$ contains no equation, then $\cI$ is  the \emph{Least Herbrand Model} of $\cS$ 
\cite{Hodges_AShorterModelTheory_1997}.
A \emph{positive boolean combination of atoms} is a formula 
\begin{eqnarray}
\bigvee_{i=1}^m\bigwedge_{j=1}^{n_i} A_{ij}\label{PositiveBooleanCombination}
\end{eqnarray}
where $m\geq 0$, $n_i\geq 0$ for all $1\leq i\leq m$, and $A_{ij}$ are atoms for all $1\leq i\leq m$ and $1\leq j\leq n_i$
(cf.\ \cite[Section 2.4]{Hodges_AShorterModelTheory_1997}).
Satisfiability of the \emph{existential closure} of formulas (\ref{PositiveBooleanCombination}), i.e., formulas of the 
form
\begin{eqnarray}
(\exists x_1)\cdots(\exists x_k) \bigvee_{i=1}^m\bigwedge_{j=1}^{n_i} A_{ij}\label{ExistentialClosureOfPositiveBooleanCombination}
\end{eqnarray} 
for some atoms $A_{ij}$ with variables in $x_1,\ldots,x_k$ for some $k\geq 0$, is \emph{preserved} under homomorphism, i.e.,
the following holds:

\begin{theorem}{\rm  \cite[cf.\ Theorem 2.4.3(a)]{Hodges_AShorterModelTheory_1997}}
\label{TheoInitialModelAndValidityOfExistentialClosure}
 Let $\SPSignature$ be a signature with predicates and $A_{ij}$ be atoms for all $1\leq i\leq m$ and $1\leq j\leq n_i$ with
 variables $x_1,\ldots,x_k$. Let $\SStructure$ and $\SStructure'$ be $\SPSignature$-structures such that 
 there is  an $\SPSignature$-homomorphism from $\SStructure$ to $\SStructure'$. Then,
 \begin{eqnarray}
\SStructure\models(\exists x_1)\cdots(\exists x_k)\bigvee_{i=1}^m\bigwedge_{j=1}^{n_i} A_{ij} & \Longrightarrow & \SStructure'\models(\exists x_1)\cdots(\exists x_k)\bigvee_{i=1}^m\bigwedge_{j=1}^{n_i} A_{ij}~~~~\label{ManyQuantifiedClause} 
\end{eqnarray}
\end{theorem}
Our main result is just a combination of the two previous results.
If $\cS$ is (logically equivalent to) a set of ground atoms, then it is satisfiable in the initial model $\InitialModel_\cS$ of $\cS$  (i.e.,
$\InitialModel_\cS\models\cS$ holds) and for all models $\SStructure$ of $\cS$ there is a homomorphism 
$h:\InitialModel_\cS\to\SStructure$ (Theorem \ref{TheoExistenceOfInitialModel}).
By Theorem \ref{TheoInitialModelAndValidityOfExistentialClosure}, if $\InitialModel_\cS$  satisfies a formula $\varphi$ of the form (\ref{ExistentialClosureOfPositiveBooleanCombination}), then for all such models $\SStructure$ of $\cS$ (for which we have a
homomorphism $h:\InitialModel_\cS\to\SStructure$) we have 
$\SStructure\models\varphi$. Thus, $\varphi$ is a logical consequence of $\cS$: $\cS\models\varphi$.

\begin{corollary}
\label{CoroInitialModelAndValidityOfExistentialClosure}
 Let $\SPSignature$ be a first-order signature, $\cS$ be a set of ground atoms, and $A_{ij}$ be atoms for all $1\leq i\leq m$ and $1\leq j\leq n_i$ with
 variables $x_1,\ldots,x_k$. Then,
\begin{eqnarray}
\InitialModel_\cS\models(\exists x_1)\cdots(\exists x_k)\bigvee_{i=1}^m\bigwedge_{j=1}^{n_i} A_{ij} & \Longrightarrow & \cS\models(\exists x_1)\cdots(\exists x_k)\bigvee_{i=1}^m\bigwedge_{j=1}^{n_i} A_{ij}~~~~\label{ManyQuantifiedClause} 
\end{eqnarray}
\end{corollary}
Corollary \ref{CoroInitialModelAndValidityOfExistentialClosure} does not hold for universally quantified formulas
or when negated atoms are present (stronger requirements on the homomorphisms are required, see  \cite[Theorems 2.4.1 and 2.4.3(b,c)]{Hodges_AShorterModelTheory_1997}).

\begin{example}
Let $\cS=\{\FP(\Fa)\}$ and $\varphi=(\forall x)\FP(x)$, which clearly holds in the least Herbrand model of $\cS$. 
The structure $\SStructure$ with domain $\naturals$ that interprets $\Fa$ as $0$ and $\FP$ as $\{0\}$ is a 
model of $\cS$ but $\SStructure\models\varphi$ does not hold.
Thus, $\cS\models\varphi$ does \emph{not} hold.

Add a new constant symbol $\Fb$ to the previous signature and consider $\varphi'=(\exists x)\neg\FP(x)$. 
Clearly, $\cI_\cS\models\varphi'$ holds.
The structure $\SStructure'$ over $\{0\}$, interpreting both $\Fa$ and $\Fb$ as $0$ and
$\FP$ again as $\{0\}$, is a model of $\cS$, but  $\SStructure'\models\varphi'$ does \emph{not} hold. 
\end{example}

\begin{remark}[Application to Horn theories]
If the (possibly infinite) set of atoms $\cS$ is viewed as generated by a finite subset $\cS_0$ of (non-necessarily atomic) Horn sentences,
then the interpretation of each predicate symbol $P$ by $\cI$ consists of the set of \emph{atomic consequences} of the form $P(t_1,\ldots,t_n)$
of $\cS$ for ground terms $t_1,\ldots,t_n$, i.e., the set of ground atoms $P(t_1,\ldots,t_n)$ such that $\cS_0\vdash P(t_1,\ldots,t_n)$
\cite{EmdKow_TheSemanticsOfPredicateLogicAsAProgrammingLanguage_JACM76}.
In order to obtain a non-empty set of ground atoms associated to a Horn theory $\cS_0$, the set of ground terms cannot be empty, i.e.,
the signature must contain at least a constant symbol.
\end{remark}
The following consequence of Corollary \ref{CoroInitialModelAndValidityOfExistentialClosure} is the basis of the
practical applications discussed in the following sections.

\begin{corollary}[Semantic criterion]\label{CoroSemanticApproach}
Let $\cS$ be a Horn theory with a non-empty set of ground atomic consequences, 
$\varphi$ be the existential closure of a positive boolean combination of atoms, 
and $\SModel$ be a model of $\cS$, i.e., $\SModel\models\cS$.
If $\SModel\models\neg\varphi$, then $\InitialModel_\cS\models\neg\varphi$.
\end{corollary}
Models $\SStructure$ to be used in Corollary \ref{CoroSemanticApproach} can be automatically generated from
the Horn theory $\cS$ and sentence $\varphi$ at stake  by using 
a tool like \AGES\ \cite{GutLucRei_AToolForTheAutomaticGenerationOfLogicalModelsForOrderSortedFirstOrderTheories_PROLE16}.
Actually, we generate a model $\SStructure$ of $\cS\cup\{\neg\varphi\}$ as described in \cite{LucGut_AutomaticSynthesisOfLogicalModelsForOrderSortedFirstOrderTheories_JAR17}.
Corollaries \ref{CoroInitialModelAndValidityOfExistentialClosure} and \ref{CoroSemanticApproach} easily generalize 
to many-sorted signatures: as usual (see \cite{Wang_LogicOfManySortedTheories_JSL52}), 
we just need to treat sorted variables $x_i:s_i$ using atoms $S_i(x_i)$ which are added as a new 
conjunction $\bigwedge_{i=1}^k S_i(x_i)$ to the matrix formula (\ref{PositiveBooleanCombination}).
In Section \ref{SecRunnningExampleSecureWeb} we use this without further formalization (but see 
\cite{GogMes_ModelsAndEqualityForLogicalProgramming_TAPSOFT87}).

\section{Conditional Rewrite Systems as Horn Theories}\label{SecCTRSsAsHornTheories}

A CTRS is a pair $\cR=(\Symbols,R)$ where $\Symbols$ is a signature of function symbols and $R$ is a set of conditional rules
$\ell\to r\IF c$ where $\ell$ and $r$ are terms and $c$ is the \emph{conditional part} of the rule
consisting of sequences $s_1\approx t_1,\ldots,s_n\approx t_n$ of expressions 
$s_i\approx t_i$, usually interpreted as \emph{reachability} or \emph{joinability} problems after an appropriate instantiation 
with a substitution $\sigma$, 
i.e., for all $i$, $1\leq i\leq n$, $\sigma(s_i)\to^*_\cR\sigma(t_i)$
(for the \emph{rewriting semantics}); or $\sigma(s_i)\downarrow_\cR\sigma(t_i)$ (for the joinability semantics)
\cite{BerKlo_ConditionalRewriteRulesConfluenceAndTermination_JCSS86,%
DerOka_RationaleForConditionalEqProgramming_TCS90,%
Ohlebusch_AdvTopicsTermRew_2002}.
In the following we focus on the reachability semantics for CTRSs\footnote{Note that the joinability semantics can be rephrased
into a reachability semantics: a joinability condition $s\downarrow t$ is equivalent to a reachability condition $s\to^*x,t\to^*x$ if $x$ is
a fresh variable not occurring elsewhere in the rule.}.
\begin{figure}[t]
\footnotesize
\begin{tabular}{|@{\quad}l@{\quad}c@{\qquad}l@{\quad}c@{\quad}|}
\hline&&&\\
(Rf) & $\bigfrac{}{x \rightarrow^{\ast} x}$  & (C) & $\bigfrac{x_i \rightarrow y_i}
{f(x_{1},\ldots,x_{i},\ldots,x_{k}){}\rightarrow{}f(x_{1},\ldots,y_{i},\ldots,x_{k})}$ 
\\& & & for all $f\in\cF$ and $1\leq i\leq k=ar(f)$
\\&&&\\
(T) & $\bigfrac{x \rightarrow z\qquad z
  \rightarrow^{*} y}{x  
\rightarrow^{*} y}$ &  (Rp) &
$\bigfrac{s_1\to^*t_1  \; \; \cdots \;\; s_n\to^*t_n }{
\ell  \rightarrow r }
$
\\ & && for $\ell \rightarrow r \IF s_1\to t_1, \ldots, s_n\to t_n\in\cR$
\\[0.3cm]
\hline
\end{tabular}
\caption{Inference rules for conditional rewriting with a CTRS $\cR$ with signature $\Symbols$}
\label{fig:inference}
\end{figure}
We write $s\to^*_\cR t$ for terms $s$ and $t$ iff there is a proof tree for 
$s\to^* t$ using $\cR$ in the inference system of Figure \ref{fig:inference}
where each rewriting step $s\to_\cR t$ also requires a \emph{proof} of the goal $s\to t$
before it can be considered part of the one-step rewriting relation associated to $\cR$ (see Figure \ref{fig:inference})
\cite{LucMarMes_OpTermCTRSs_IPL05}.

\begin{remark}
All rules in the inference system in Figure \ref{fig:inference} 
are \emph{schematic} in the sense that each inference rule $\frac{B_1~\cdots~B_n}{A}$
can be used for any \emph{instance} $\frac{\sigma(B_1)~\cdots~\sigma(B_n)}{\sigma(A)}$
of the rule by a substitution $\sigma$ \cite{Smullyan_TheoryOfFormalSystems_1961}. 
For instance, (Rp) actually establishes that, for every rule $\ell\to r\IF s_1\to t_1,\ldots,s_n\to t_n$
in the CTRS $\cR$, every instance $\sigma(\ell)$ by a substitution
$\sigma$ rewrites into $\sigma(r)$ provided that, for each $s_i\to t_i$, with $1\leq i\leq n$, the reachability condition $\sigma(s_i)\to^*\sigma(t_i)$ can be \emph{proved}.
\end{remark}
In the logic of CTRSs, with binary \emph{predicates} $\to$ and $\to^*$, the \emph{Horn theory} $\ol{\cR}$ for a CTRS $\cR$ 
 is obtained from the inference rules in Figure \ref{fig:inference} (for the \emph{reachability semantics of conditions})
 by 
\emph{specializing} 
$(\mathit{C})_{f,i}$ for each $f\in\Symbols$ and $i$, $1\leq i\leq ar(f)$
and $(\mathit{Rp})_\rho$ for all $\rho:\ell\to r\IF c\in R$.
Inference rules $\frac{B_1~\cdots~B_n}{A}$ become universally quantified \emph{implications} 
$B_1\wedge\cdots\wedge B_n\Rightarrow A$ 
\cite[Section 2]{LucMes_ModelsForLogicsAndConditionalConstraintsInAutomatedProofsOfTermination_AISC14}.
\begin{example}\label{Exhosc08_ex5_aecc_p46}
For the following CTRS $\cR$ \cite[page 46]{GieArt_VerificationErlangProcessesDPs_AAECC01}:\\[-0.3cm]
\begin{tabular}{cc}
\begin{minipage}{5.1cm}
\begin{eqnarray}
   \Fa & \to & \Fb\label{Exhosc08_ex5_aecc_p46_rule1}\\
\Ff(\Fa) & \to & \Fb\label{Exhosc08_ex5_aecc_p46_ule2}
\end{eqnarray}
\end{minipage}
&
\begin{minipage}{7cm}
\begin{eqnarray}
\Fg(x) & \to & \Fg(\Fa) \IF \Ff(x) \to x\label{Exhosc08_ex5_aecc_p46_rule3}\\
&&\nonumber
\end{eqnarray}
\end{minipage}
\end{tabular}
\smallskip

\noindent
Figure \ref{FigExhosc08_ex5_aecc_p46_FOtheory} shows its Horn theory  $\ol{\cR}$.
\begin{figure}[t]
\begin{eqnarray}
& (\forall x)~x \to^* x\label{Exhosc08_ex5_aecc_p46_sentenceRefl}\\
& (\forall x,y,z)~(x\to y\wedge y \to^* z\Rightarrow x\to^* z)\label{Exhosc08_ex5_aecc_p46_sentenceTran}\\
& (\forall x,y)~(x\to y \Rightarrow \Ff(x)\to \Ff(y))\label{Exhosc08_ex5_aecc_p46_sentenceCf1}\\
& (\forall x,y)~(x\to y \Rightarrow \Fg(x)\to \Fg(y))\label{Exhosc08_ex5_aecc_p46_sentenceCg1}\\
&\Fa \to \Fb\label{Exhosc08_ex5_aecc_p46_sentenceRepl1}\\
&\Ff(\Fa) \to \Fb\label{Exhosc08_ex5_aecc_p46_sentenceRepl2}\\
& (\forall x)\:(\Ff(x) \to^* x \Rightarrow \Fg(x) \to \Fg(\Fa))\label{Exhosc08_ex5_aecc_p46_sentenceRepl3}
\end{eqnarray}
\caption{Horn theory for $\cR$ in Example \ref{Exhosc08_ex5_aecc_p46}}
\label{FigExhosc08_ex5_aecc_p46_FOtheory}
\end{figure}
\end{example}

\section{Application to (Conditional) Term Tewriting}\label{SecApplicatonToRewriting}

Note that all sentences 
in Figure \ref{FigSomeFOpropertiesRewritingBasedSystems}
are particular cases of (\ref{ExistentialClosureOfPositiveBooleanCombination}) when the language of the logic of
CTRSs is used.
Some of the problems represented by these formulas have been investigated in the literature.
In the following, we consider them and show that our results are useful to improve or complement
the already developed proof methods for these analysis problems.

\subsection{Infeasible Conditional Critical Pairs ($\varphi_{\mathit{Feas}}$)}\label{SecInfeasibleCCPs}

In the literature about \emph{confluence} of conditional rewriting, the so-called \emph{infeasible} Conditional 
Critical Pairs (CCPs) for a CTRS $\cR$ are those critical pairs $s\downarrow t\IF c$ 
whose \emph{conditional parts} $c$
are \emph{infeasible}, i.e., there is no substitution $\sigma$ such that for all $i$, $1\leq i\leq n$,
we have $\sigma(s_i)\to^*_\cR\sigma(t_i)$
(for the \emph{rewriting semantics}; or $\sigma(s_i)\downarrow_\cR\sigma(t_i)$ for the joinability semantics)
\cite[Definition 7.1.8]{Ohlebusch_AdvTopicsTermRew_2002}. 
Detecting infeasible CCPs is important in proofs of confluence of CTRSs 
\cite{BerKlo_ConditionalRewriteRulesConfluenceAndTermination_JCSS86,Ohlebusch_AdvTopicsTermRew_2002,%
SteMid_InfeasibleConditionalCriticalPairs_IWC15,SteSte_CertifyingConfluenceOfAlmostOrthogonalCTRSsViaExactTreeAutomataCompletion_FSCD16}.

Although infeasibility of CCPs is undecidable, recent tools developed to prove confluence of CTRSs (e.g., \cite{SteMid_ConditionalConfluenceSystemDescription_RTATLCA14})
implement a number of  sufficient criteria to prove infeasibility of CCPs
\cite{SteMid_InfeasibleConditionalCriticalPairs_IWC15,SteSte_CertifyingConfluenceOfAlmostOrthogonalCTRSsViaExactTreeAutomataCompletion_FSCD16}.
Infeasibility of CCPs with respect to a CTRS $\cR$
can be investigated using $\varphi_{\mathit{Feas}}$, i.e.,
$(\exists \vec{x})\:s_1\to^*t_1\wedge\cdots\wedge s_n\to^*t_n$ 
(see Figure \ref{FigSomeFOpropertiesRewritingBasedSystems}) together with 
Corollary \ref{CoroSemanticApproach}.

\begin{example}\label{Ex5_1_SM15}
The following CTRS \cite[Example 5.1]{SteMid_InfeasibleConditionalCriticalPairs_IWC15}
\[
\begin{array}{r@{~}c@{~}l@{\hspace{0.4cm}}r@{~}c@{~}l@{\hspace{0.4cm}}r@{~}c@{~}l@{\hspace{0.4cm}}r@{~}c@{~}l@{\hspace{0.4cm}}}
\Fz \leq x & \to & \Ftrue  & \Fs(x) > \Fz & \to & \Ftrue & x - \Fz & \to & x\\
\Fs(x) \leq \Fs(y) & \to & x\leq y  & \Fs(x) > \Fs(y) & \to & x> y & \Fz-x & \to &\Fz & \Fs(x)-\Fs(y) & \to & x-y
\end{array}
\]
\[\begin{array}{r@{~}c@{~}l@{\hspace{0.5cm}}r@{~}c@{~}l@{\hspace{0.5cm}}r@{~}c@{~}l@{\hspace{0.5cm}}r@{~}c@{~}l@{\hspace{0.5cm}}}
x \div y\to \langle 0,y\rangle & \IF & y>x\to\Ftrue\\
x \div y\to \langle\Fs(q),r\rangle & \IF & y\leq x\to\Ftrue, (x-y)\div x\to\langle y,z\rangle\\
\end{array}
\]
has the following conditional critical pair:
\begin{eqnarray}
\langle\Fz,x\rangle\downarrow\langle\Fs(y),z\rangle\IF x\leq w\to\Ftrue, (w-x)\div x\to\langle y,z\rangle,x>w\to\Ftrue\nonumber\end{eqnarray}
The structure $\SStructure$ below\footnote{All models displayed in the examples of this paper have been computed with \AGES.}
provides a model of $\ol{\cR}\cup\{\neg\varphi_{\mathit{Feas}}\}$ where $\varphi_{\mathit{Feas}}$ is 
\begin{eqnarray}
(\exists w,x,y,z)\:(x\leq w\to^*\Ftrue, (w-x)\div x\to^*\langle y,z\rangle,x>w\to^*\Ftrue)\label{Ex5_1_SM15_CCPcondition}
\end{eqnarray}
The domain of $\SemDomain$ is the set of natural numbers $\naturals$. Function symbols are interpreted as follows:
\[
\begin{array}{r@{~}c@{~}l@{\hspace{0.5cm}}r@{~}c@{~}l@{\hspace{0.5cm}}r@{~}c@{~}l@{\hspace{0.5cm}}r@{~}c@{~}l@{\hspace{0.5cm}}}
\Ftrue^\SStructure & = & 1 & \Fz^\SStructure & = & 0 & \Fs^\SStructure(x) & = & x+1 \\
x\leq^\SStructure y & = & 
\left \{
\begin{array}{cl}
1 & \text{if }y\geq_\naturals x\\
0 & \text{otherwise}
\end{array}
\right .
&
x>^\SStructure y & = & 
\left \{
\begin{array}{cl}
1 & \text{if }x >_\naturals y\\
0 & \text{otherwise}
\end{array}
\right .
&
x\div^\SStructure y & = & 1\\[0.3cm]
x-^\SStructure y & = & 
\left \{
\begin{array}{cl}
x-_\naturals y & \text{if }x \geq_\naturals y\\
0 & \text{otherwise}
\end{array}
\right .
&
\langle x,y\rangle^\SStructure & = & 1\\

\end{array}
\]
Predicate symbols $\to$ and $\to^*$ are interpreted as follows:
\[
\begin{array}{r@{~}c@{~}l@{\hspace{0.5cm}}r@{~}c@{~}l@{\hspace{0.5cm}}r@{~}c@{~}l@{\hspace{0.5cm}}r@{~}c@{~}l@{\hspace{0.5cm}}}
x\to y & \Leftrightarrow & x=_\naturals y& x\to^* y & \Leftrightarrow & x\geq_\naturals y
\end{array}
\]
Thus, the critical pair is infeasible.
In  \cite[Example 5.1]{SteMid_InfeasibleConditionalCriticalPairs_IWC15} this is
proved by using the theorem prover {\sf Waldmeister} \cite{GaiHilLocSpi_TheNewWaldmeisterLoopAtWork_CADE03}.
\end{example}

\begin{example}\label{Ex23_SS16}
The following CTRS $\cR$ \cite[Example 23]{SteSte_CertifyingConfluenceOfAlmostOrthogonalCTRSsViaExactTreeAutomataCompletion_FSCD16}
\begin{eqnarray}
\Fg(x) & \to & \Ff(x,x)\label{Ex23_SS16_rule1}\\
\Fg(x) & \to & \Fg(x) \IF \Fg(x) \to \Ff(\Fa,\Fb)\label{Ex23_SS16_rule2}
\end{eqnarray}
has a conditional critical pair 
$\Ff(x,x) \downarrow \Fg(x)\IF \Fg(x)\to\Ff(\Fa,\Fb)$. 
The following structure $\SStructure$ over the finite domain $\{0,1\}$:
\[\begin{array}{r@{\:}c@{\:}l@{\hspace{0.6cm}}r@{\:}c@{\:}l@{\hspace{0.6cm}}r@{\:}c@{\:}l@{\hspace{0.6cm}}r@{\:}c@{\:}l}
\Fa^\SStructure & = & 1 & \Fb^\SStructure = \Fc^\SStructure & = & 0 & 
\Ff^\SStructure(x,y) & = & 
\left \{
\begin{array}{cl}
x - y + 1 & \text{if } x\geq y\\
y - x + 1 & \text{otherwise}
\end{array}
\right . \\
\Fg^\SStructure(x) & = & 1 &
x \to^\SStructure y & \Leftrightarrow & x= y  & x\:(\to^*)^\SStructure\:y & \Leftrightarrow & x\geq y
\end{array}
\]
is a model $\ol{\cR}\cup\{\neg\varphi_{\mathit{Feas}}\}$ for $\varphi_{\mathit{Feas}}$ given by $(\exists x)~\Fg(x)\to^*\Ff(\Fa,\Fb)$.
Thus, the critical pair is infeasible. 
In \cite[Example 23]{SteSte_CertifyingConfluenceOfAlmostOrthogonalCTRSsViaExactTreeAutomataCompletion_FSCD16} this is
proved by using unification tests together with a transformation. 
It is discussed that the the alternative tree automata techniques investigated in the paper do \emph{not} work for this example.
\end{example}

\subsection{Infeasible Rules ($\varphi_{\mathit{Feas}}$)}\label{SecProcInfeasibleRules}

The infeasibility of the \emph{conditional part} of a conditional rule with respect to a given CTRS
is also important to prove other computational properties of such systems. In particular, proving
the infeasibility of the \emph{conditional dependency pairs} which are used to characterize termination properties
of CTRSs \cite{LucMes_DependencyPairsForProvingTerminationPropertiesOfCTRSs_JLAMP17}
is useful in (automated) proofs of such termination properties \cite{LucMesGut_ExtendingThe2DDPFrameworkForCTRSs_LOPSTR14}.

\begin{example}
\label{Exhosc08_ex5_aecc_p46_InfeasibleRuleSemanticCriterion}
A CTRS $\cR$ is \emph{operationally terminating} iff no term $t$ has
an infinite proof tree using the inference system in Figure \ref{fig:inference} \cite{LucMarMes_OpTermCTRSs_IPL05}.
According to \cite{LucMes_DependencyPairsForProvingTerminationPropertiesOfCTRSs_JLAMP17,LucMesGut_ExtendingThe2DDPFrameworkForCTRSs_LOPSTR14}, 
a formal proof of operational termination of $\cR$ in Example \ref{Exhosc08_ex5_aecc_p46} is easily obtained 
if the following \emph{conditional dependency pair} (which is just a conditional rule):
\begin{eqnarray}
\FG(x) & \to & \FG(\Fa) \IF \Ff(x) \to x\label{Exhosc08_ex5_aecc_p46_DPH1}
\end{eqnarray}
(where $\FG$ is a new function symbol)
is proved \emph{infeasible} with respect to reductions with $\cR$.
The following structure $\SStructure$ over $\naturals-\{0\}$:
\[\begin{array}{r@{\:}c@{\:}l@{\hspace{0.6cm}}r@{\:}c@{\:}l@{\hspace{0.6cm}}r@{\:}c@{\:}l@{\hspace{0.6cm}}r@{\:}c@{\:}l}
\Fa^\SStructure & = & 1 & \Fb^\SStructure & = & 2 & 
\Ff^\SStructure(x) & = & x+1 & \Fg^\SStructure(x) & = & 1 \\
x \to^\SStructure y & \Leftrightarrow & x\leq y  & x\:(\to^*)^\SStructure\:y & \Leftrightarrow & x\leq y 
\end{array}
\]
is a model of $\ol{\cR}\cup\{\neg\varphi_{\mathit{Feas}}\}$, where $\ol{\cR}$ is in 
Figure \ref{FigExhosc08_ex5_aecc_p46_FOtheory} and $\varphi_{\mathit{Feas}}$ is $(\exists x)~\Ff(x) \to^* x$.
Thus, rule
$(\ref{Exhosc08_ex5_aecc_p46_DPH1})$
is proved $\cR$-infeasible and $\cR$ operationally terminating.
\end{example}
\begin{example}\label{Ex17_SS16}
Consider the following CTRS $\cR$ \cite[Example 17]{SteSte_CertifyingConfluenceOfAlmostOrthogonalCTRSsViaExactTreeAutomataCompletion_FSCD16}:\\[-0.8cm]

\begin{tabular}{cc}
\hspace{-1cm}
\begin{minipage}[t]{.5\linewidth}
\begin{eqnarray}
\Fh(x) & \to & \Fa\label{Ex17_SS16_rule1}\\
\Fg(x) & \to & x\label{Ex17_SS16_rule2}
\end{eqnarray}\nonumber
\end{minipage} & 
\begin{minipage}[t]{.52\linewidth}
\begin{eqnarray}
\Fg(x) & \to & \Fa \IF \Fh(x) \to \Fb\label{Ex17_SS16_rule3}\\
\Fc & \to & \Fc\label{Ex17_SS16_rule4}
\end{eqnarray}\nonumber
\end{minipage}
\end{tabular}\\[0.3cm]
The following structure $\SStructure$ over $\naturals$:
\[\begin{array}{r@{\:}c@{\:}l@{\hspace{0.6cm}}r@{\:}c@{\:}l@{\hspace{0.6cm}}r@{\:}c@{\:}l@{\hspace{0.6cm}}r@{\:}c@{\:}l}
\Fa^\SStructure & = & 0 & \Fb^\SStructure = \Fc^\SStructure & = & 1 & 
\Fg^\SStructure(x) & = & x+2 & \Fh^\SStructure(x) & = & 0 \\
x \to^\SStructure y & \Leftrightarrow & x\geq y  & x\:(\to^*)^\SStructure\:y & \Leftrightarrow & x\geq y
\end{array}
\]
is a model of $\ol{\cR}\cup\{\neg\varphi_{\mathit{Feas}}\}$ where $\varphi_{\mathit{Feas}}$ is $((\exists x)~\Fh(x) \to^* \Fb)$.
Therefore, rule 
$(\ref{Ex17_SS16_rule3})$
is proved $\cR$-infeasible. 
In \cite[Example 17]{SteSte_CertifyingConfluenceOfAlmostOrthogonalCTRSsViaExactTreeAutomataCompletion_FSCD16} this is
proved by using tree automata techniques.
It is also shown that the alternative technique investigated in the paper (the use of unification tests) does \emph{not} work in this case.
\end{example}

\subsection{Non-Joinability of Critical Pairs ($\varphi_{\mathit{Join}}$)}

The analysis of \emph{confluence} often
relies on checking for \emph{joinability} of the components $s$ and $t$ of a \emph{critical pair} $s\downarrow t$ obtained from the
rules of the (C)TRS $\cR$, i.e., we look for a term $u$ such that $s\to^*_\cR u$ and $t\to^*_\cR u$.
The problem of \emph{disproving} joinability of \emph{ground} terms has been 
investigated for TRSs, as an interesting contribution to the
development of methods for (automatically) proving \emph{non-confluence} of TRSs \cite{Aoto_DisprovingConfluenceOfTermRewritingSystemsByInterpretationAndOrdering_FroCoS13}.

Actually, proving \emph{non-joinability} of (ground) terms can be seen as a particular case of \emph{infeasibility}: given
ground terms $s$ and $t$, we
prove that $(\exists x)\:(s\to^*x \wedge t\to^*x)$ does \emph{not} hold.
In this way, we use our technique to check non-joinability of ground terms in CTRSs, something which is
also considered in \cite{SteSte_CertifyingConfluenceOfAlmostOrthogonalCTRSsViaExactTreeAutomataCompletion_FSCD16}.

\begin{example}\label{Ex3_SS16}
The following CTRS $\cR$ \cite[Example 3]{SteSte_CertifyingConfluenceOfAlmostOrthogonalCTRSsViaExactTreeAutomataCompletion_FSCD16}
\begin{eqnarray}
\Ff(x) & \to & \Fa \IF x\to \Fa\label{Ex3_SS16_rule1}\\
\Ff(x) & \to & \Fb \IF x\to \Fb\label{Ex3_SS16_rule2}
\end{eqnarray}
has a conditional critical pair 
$\Fa \downarrow \Fb\IF x\to\Fa, x\to\Fb$. 
This critical pair is both \emph{non-joinable} and  \emph{infeasible}: 
\begin{enumerate}
\item For non-joinability, consider the structure $\SStructure$ over  $\{0,1\}$:
\[\begin{array}{r@{\:}c@{\:}l@{\hspace{0.6cm}}r@{\:}c@{\:}l@{\hspace{0.6cm}}r@{\:}c@{\:}l@{\hspace{0.6cm}}r@{\:}c@{\:}l}
\Fa^\SStructure & = & 0 & \Fb^\SStructure  & = & 1 & 
\Ff^\SStructure(x) & = & x\\
x \to^\SStructure y & \Leftrightarrow & x= y  & x\:(\to^*)^\SStructure\:y & \Leftrightarrow & x= y
\end{array}
\]
which is a model $\ol{\cR}\cup\{\neg\varphi_{\mathit{Join}}\}$ for $\varphi_{\mathit{Join}}$ given by $(\exists x)~\Fa\to^*x\wedge \Fb\to^*x$.
Thus, the critical pair is non-joinable. 
In \cite[Example 3]{SteSte_CertifyingConfluenceOfAlmostOrthogonalCTRSsViaExactTreeAutomataCompletion_FSCD16} this is
proved by an unification test. 
\item For infeasibility, consider the structure $\SStructure$ over  $\naturals$:
\[\begin{array}{r@{\:}c@{\:}l@{\hspace{0.6cm}}r@{\:}c@{\:}l@{\hspace{0.6cm}}r@{\:}c@{\:}l@{\hspace{0.6cm}}r@{\:}c@{\:}l}
\Fa^\SStructure & = & 1 & \Fb^\SStructure  & = & 0 & 
\Ff^\SStructure(x) & = & x\\
x \to^\SStructure y & \Leftrightarrow & x= y  & x\:(\to^*)^\SStructure\:y & \Leftrightarrow & x= y
\end{array}
\]
which is a model $\ol{\cR}\cup\{\neg\varphi_{\mathit{Feas}}\}$ for $\varphi_{\mathit{Feas}}$ given by $(\exists x)~x\to^*\Fa\wedge x\to^*\Fb$.
Thus, the critical pair is infeasible. 
In \cite[Example 3]{SteSte_CertifyingConfluenceOfAlmostOrthogonalCTRSsViaExactTreeAutomataCompletion_FSCD16} this is
not actually proved but the authors argue that the unification test does not work. 
\end{enumerate}
\end{example}

\subsection{Irreducible Terms ($\varphi_{\mathit{Red}}$)}

It is well-known that, in sharp contrast to unconditional rewriting, for CTRSs $\cR$ it is \emph{not} decidable whether a given term
$t$ is (one-step) reducible.
In Example \ref{ExLoopingnessInCTRSs_ProofOfIrreducible},  we already exemplified the use of our technique to check whether a given reduction step $s\to t$ for ground terms $s$ and $t$ 
is \emph{not} possible.
In general, with
$\varphi_{\mathit{Red}}$, i.e., $(\exists x)\:t\to x$, and Corollary \ref{CoroSemanticApproach} we can prove that a given ground term $t$ is
\emph{irreducible}.
In the following example we show an interesting variant.

\begin{example}\label{Ex13_LM16}
Consider the following CTRS $\cR$ \cite[Example 13]{LucMes_NormalFormsAndNormalTheoriesunconditionalRewriting_JLAMP16}:\\[-0.6cm] 
\begin{tabular}{c@{\hspace{-0.6cm}}c}
\hspace{-1cm}
\begin{minipage}[t]{.58\linewidth}
\begin{eqnarray}
\Fa & \to & \Fb\\
\Fb & \to & \Fa
\end{eqnarray}\nonumber
\end{minipage} & 
\begin{minipage}[t]{.55\linewidth}
\begin{eqnarray}
\Ff(x) & \to & x \IF \Fc\to\Fd, \Fa\to\Fc
\end{eqnarray}\nonumber
\end{minipage}
\end{tabular}\\[0.3cm]
Note that every term $\Ff(t)$
is \emph{irreducible at the root}. We can prove this claim with a slight variant of $\varphi_{\mathit{Red}}$: $(\exists x,y)~\Ff(x)\stackrel{\toppos}{\to} y$, which claims
for the existence of a \emph{root-reducible} instance $\Ff(t)$ of $\Ff(x)$.
The new predicate $\stackrel{\toppos}{\to}$ has a slightly different Horn theory $H_{\stackrel{\toppos}{\to}_\cR}$ where reductions with 
$\stackrel{\toppos}{\to}_\cR$ are \emph{not} propagated below
the root of terms: for each rule $\ell\to r\IF s_1\to t_1,\ldots,s_n\to t_n$, we have a sentence:
\begin{eqnarray}
(\forall x_1,\ldots,x_k)~s_1\to^*t_1\wedge\cdots\wedge s_n\to^*t_n\Rightarrow \ell &  \stackrel{\toppos}{\to} & r\label{RootRule}
\end{eqnarray}
in $H_{\stackrel{\toppos}{\to}_\cR}$ (where $x_1,\ldots,x_k$ are the variables occurring in the rule) and nothing else.
Note that the conditions in the rules are evaluated with $\to^*_\cR$ rather than with $\toStarPosWith{\toppos}{\cR}$.
For this reason, no definition of the reflexive and transitive closure of $\stackrel{\toppos}{\to}$ is given.
Thus, the Horn theory $\ol{\cR}\cup H_{\stackrel{\toppos}{\to}_\cR}$ we have to deal with is\\[-0.6cm] 
\begin{tabular}{c@{\hspace{-0.6cm}}c}
\hspace{-1cm}
\begin{minipage}[t]{.58\linewidth}
\begin{eqnarray}
& (\forall x)\:x \to^* x\label{Ex13_LM16_sentenceRefl}\\
& (\forall x,y,z)\:(x\to y\wedge y \to^* z\Rightarrow x\to^* z)\label{Ex13_LM16_sentenceTran}\\
& (\forall x,y)\:(x\to y \Rightarrow \Ff(x)\to \Ff(y))\label{Ex13_LM16_sentenceCc1}\\
&\Fa \to \Fb\label{Ex13_LM16_sentenceRepl1}\\
&\Fb \to \Fa\label{Ex13_LM16_sentenceRepl2}\\
& (\forall x)\:\Fc\to^*\Fd\wedge \Fa\to^*\Fc\Rightarrow \Ff(x) \to x \label{Ex13_LM16_sentenceRepl3}
\end{eqnarray}\nonumber
\end{minipage} & 
\begin{minipage}[t]{.55\linewidth}
\begin{eqnarray}
& \Fa \stackrel{\toppos}{\to} \Fb\label{Ex13_LM16_sentenceRootRed1}\\
&\Fb \stackrel{\toppos}{\to} \Fa\label{Ex13_LM16_sentenceRootRed2}\\
& (\forall x)\:\Fc\to^*\Fd\wedge \Fa\to^*\Fc\Rightarrow \Ff(x) \stackrel{\toppos}{\to} x \label{Ex13_LM16_sentenceRootRed3}
\end{eqnarray}\nonumber
\end{minipage}
\end{tabular}\\[0.3cm]
with $\ol{\cR}=\{(\ref{Ex13_LM16_sentenceRefl})-(\ref{Ex13_LM16_sentenceRepl3})\}$ and $H_{\stackrel{\toppos}{\to}_\cR}=\{(\ref{Ex13_LM16_sentenceRootRed1})-(\ref{Ex13_LM16_sentenceRootRed3})\}$. The following structure $\SStructure$ over $\{-1,0,1\}$ is a 
model of $\ol{\cR}\cup H_{\stackrel{\toppos}{\to}_\cR}\cup\{\neg\varphi_{\mathit{RRed}}\}$ where $\varphi_{\mathit{RRed}}$ is 
$(\exists x,y)~\Ff(x)\stackrel{\toppos}{\to} y$:
\[\begin{array}{r@{\:}c@{\:}l@{\hspace{0.4cm}}r@{\:}c@{\:}l@{\hspace{0.4cm}}r@{\:}c@{\:}l@{\hspace{0.4cm}}r@{\:}c@{\:}l}
\Fa^\SStructure = \Fb^\SStructure & = & -1 & 
\Fc^\SStructure & = & 0 & \Fd^\SStructure & = & 1 & \Ff^\SStructure(x) & = & 1  \\
x \to^\SStructure y & \Leftrightarrow & x \geq y  & x\:(\to^*)^\SStructure\:y & \Leftrightarrow & x\geq y & x (\stackrel{\toppos}{\to})^\SStructure y & \Leftrightarrow & 5x+y\leq 1
\end{array}
\]
This proves that for all ground terms $t$, $\Ff(t)$ is irreducible at the root. 
\end{example}

\subsection{Cycling/Looping Terms and Systems ($\varphi_{\mathit{Cycl}}/\varphi_{\mathit{Loop}}$)}\label{SecCyclingLoopingTerms}

A term $t$ \emph{loops} (with respect to a CTRS $\cR$) 
if there is a rewrite sequence $t=t_1\to_\cR\cdots\to_\cR t_n$ for some $n>1$ such that $t$ is a (non-necessarily 
strict) subterm of $t_n$, written $t_n\subtermeq t$ (cf., \cite[Definition 3]{Dershowitz_TerminationOfRewriting_JSC87}).
We say that a CTRS is non-looping if no term loops.
We can 
check loopingness of terms $t$ or CTRSs $\cR$ by using $\varphi_{\mathit{Loopt}}$ and 
$\varphi_{\mathit{Loop}}$ in Figure \ref{FigSomeFOpropertiesRewritingBasedSystems} together
with Corollary \ref{CoroSemanticApproach} if the considered Horn theory is
 the union of $\ol{\cR}$ and the Horn theory $H_\subtermeq$ describing the subterm relation $\subtermeq$:
\begin{eqnarray}
(\forall x)\: x  & \subtermeq &  x\\
(\forall x,y,z)~x \subtermeq y \wedge y \subtermeq z \Rightarrow x & \subtermeq & z\\
(\forall x_1,\ldots,x_k)~f(x_1,\ldots,x_k) &  \subtermeq & x_i\label{EmbeddingRule}
\end{eqnarray}
where (\ref{EmbeddingRule}) is given for each $k$-ary function symbol $f\in\Symbols$ and argument $i$, $1\leq i\leq k$.

\begin{example}\label{ExNonLoopingNonCycling}
Consider the TRS
\begin{eqnarray}
\Fa & \to & \Fc(\Fb)\\
\Fb & \to & \Fc(\Fb)
\end{eqnarray}
We can prove $\Fa$ non-looping. The Horn theory $\ol{\cR}\cup H_\subtermeq$ 
is the following\\[-0.6cm] 
\begin{tabular}{c@{\hspace{-0.6cm}}c}
\hspace{-0.8cm}
\begin{minipage}[t]{.58\linewidth}
\begin{eqnarray}
& (\forall x)~x \to^* x\label{ExNonLoopingNonCycling_sentenceRefl}\\
& (\forall x,y,z)~(x\to y\wedge y \to^* z\Rightarrow x\to^* z)\label{ExNonLoopingNonCycling_sentenceTran}\\
& (\forall x,y)~(x\to y \Rightarrow \Fc(x)\to \Fc(y))\label{ExNonLoopingNonCycling_sentenceCc1}\\
&\Fa \to \Fc(\Fb)\label{ExNonLoopingNonCycling_sentenceRepl1}\\
&\Fb \to \Fc(\Fb)\label{ExNonLoopingNonCycling_sentenceRepl2}
\end{eqnarray}\nonumber
\end{minipage} & 
\begin{minipage}[t]{.52\linewidth}
\begin{eqnarray}
& (\forall x)\: x  \subtermeq   x\label{ExNonLoopingNonCycling_sentenceSubtermRefl}\\
& (\forall x,y,z)~x \subtermeq y \wedge y \subtermeq z \Rightarrow x  \subtermeq z\label{ExNonLoopingNonCycling_sentenceSubtermTran}\\
& (\forall x)\: \Fc(x)  \subtermeq   x\label{ExNonLoopingNonCycling_sentenceSubtermProjFc1}
\end{eqnarray}\nonumber
\end{minipage}
\end{tabular}\\[0.3cm]
The following structure over $\naturals\cup\{-1\}$:
\[\begin{array}{r@{\:}c@{\:}l@{\hspace{0.6cm}}r@{\:}c@{\:}l@{\hspace{0.6cm}}r@{\:}c@{\:}l@{\hspace{0.6cm}}r@{\:}c@{\:}l}
\Fa^\SStructure & = & -1 & \Fb^\SStructure & = & 1 & 
\Fc^\SStructure(x) & = & x \\
x \to^\SStructure y & \Leftrightarrow & x \leq 1\wedge y\geq 1  & x\:(\to^*)^\SStructure\:y & \Leftrightarrow & x\leq y  & x \subtermeq^\SStructure y & \Leftrightarrow & x\leq y
\end{array}
\]
 is a model of $\ol{\cR}\cup H_\subtermeq\cup\{\neg\varphi_{\mathit{Loopt}}\}$ where 
$\varphi_{\mathit{Loopt}}$ is $(\exists x,y)~\Fa\to x\wedge x\to^*y\wedge y\subtermeq\Fa$.
Therefore, $\Fa$ is non-looping.
On the other hand, although $\Fb$ is a looping term, we can also prove that it is \emph{non-cycling}. Actually, we can prove that $\cR$
itself is non-cycling with the following structure over $\naturals\cup\{-1\}$
\[\begin{array}{r@{\:}c@{\:}l@{\hspace{0.6cm}}r@{\:}c@{\:}l@{\hspace{0.6cm}}r@{\:}c@{\:}l@{\hspace{0.6cm}}r@{\:}c@{\:}l}
\Fa^\SStructure & = & -1 & \Fb^\SStructure & = & -1 & 
\Fc^\SStructure(x) & = & 2x +2\\
x \to^\SStructure y & \Leftrightarrow & x < y  & x\:(\to^*)^\SStructure\:y & \Leftrightarrow & x\leq y
\end{array}
\]
which is a model of $\ol{\cR}\cup \{\neg\varphi_{\mathit{Cycl}}\}$ where 
$\varphi_{\mathit{Cycl}}$ is $(\exists x,y)~x\to y\wedge y\to^*x$.
\end{example}

\subsection{Secure Access to Web sites}\label{SecRunnningExampleSecureWeb}

The specification in Figure 
\ref{FigOSSpecWWV05} provides a
partial representation of the structure and
connectivity of the  site of the $1^{st}$ International Workshop on 
Automated Specification and Verification of Web Sites, WWV'05\footnote{\url{http://users.dsic.upv.es/workshops/wwv05/}}
originally considered in \cite{Lucas_RewritingBasedNavigationOfWebSites_WWV05}.
As in \cite{Lucas_RewritingBasedNavigationOfWebSites_WWV05}, 
web pages are modeled as terms $p(u)$ where $u$ represents the \emph{user}  
browsing the site. Transitions among web pages are modeled as rewrite rules.
In contrast to \cite{Lucas_RewritingBasedNavigationOfWebSites_WWV05}, we
use an order-sorted specification and the sort of $u$ is used to allow/disallow the access to some
web pages.
For this reason, the specification is given as a \Maude\ module whose syntax is hopefully self-explanatory \cite{ClavelEtAl_MaudeBook_2007}.
\begin{figure}[t]
{\footnotesize
\begin{center}
\begin{minipage}{10cm}
\begin{verbatim}
mod WWV05-WEBSITE is
  sorts EventualUser RegUser User WebPage SecureWebPage .
  subsorts RegUser EventualUser < User .
  subsorts SecureWebPage < WebPage .

  ops login register sbmlink submission wwv05 : User -> WebPage .
  op vlogin : User -> SecureWebPage .
  op submit : RegUser -> SecureWebPage .
  
  op slucas : -> RegUser .
  op smith : -> EventualUser .

  var R : RegUser .
  var U : User .
  
  rl wwv05(U) => submission(U) .
  rl submission(U) => sbmlink(U) .
  rl sbmlink(U) => login(U) .
  rl sbmlink(U) => register(U) .
  rl login(U) => vlogin(U) .
  rl vlogin(R) => submit(R) .
endm
\end{verbatim}
\end{minipage}
\end{center}}
\caption{Maude specification of part of the WWV05 web site}\label{FigOSSpecWWV05}
\end{figure}

We want to guarantee a
 \emph{secure} access to web pages: browsing is allowed for  \emph{registered} users only.
\emph{Regular} and \emph{secure} pages are terms of 
sort $\pr{WebPage}$ and $\pr{SecureWebPage}$, respectively.
 $\pr{SecureWebPage}$ is subsort of $\pr{WebPage}$.
Registered and eventual users are given sorts $\pr{RegUser}$ and $\pr{EventualUser}$,
respectively. 
Both are subsorts of $\pr{User}$.
Browsing the web site is modeled as \emph{rewriting} in the OS-TRS above.
Our goal is verifying that
\emph{no \emph{eventual} user can reach the submission page}.
Thus, we formulate the property we want to \emph{avoid}:
\begin{eqnarray}
(\exists \mathtt{u:EventualUser}) ~\mathtt{wwv05(u)}\to^*\mathtt{submit(u)}\label{ExOSSpecWWV05_ReachabilityPropertyPositivePart}
\end{eqnarray}
Indeed, this is a particular case of $\varphi_{\mathit{Feas}}$ but including information about sorts is crucial.
The following structure $\SStructure$ with (i) domains 
\[\begin{array}{r@{\:}c@{\:}l@{\hspace{0.6cm}}r@{\:}c@{\:}l@{\hspace{0.6cm}}r@{\:}c@{\:}l@{\hspace{0.6cm}}r@{\:}c@{\:}l@{\hspace{0.6cm}}r@{\:}c@{\:}l}
\SStructure_{\mathtt{EventualUser}} & = &  \{-1\} & 
\SStructure_{\mathtt{RegUser}} & = &  \{1\} & 
\SStructure_{\mathtt{User}} & = & \naturals\cup\{-1\}\\
\SStructure_{\mathtt{WebPage}} & = & \{-1\} & 
\SStructure_{\mathtt{SecureWebPage}} & = & \{-1\} 
\end{array}
\]
(ii) function symbols interpreted by 
\begin{itemize}
\item $f^\SStructure(x)=-1$ for $f\in\{\pr{login},\pr{register},\pr{sbmlink},\pr{submission},\pr{vlogin},\pr{wwv05}\}$, 
\item $\pr{submit}^\SStructure(x)=-x$,
\item $\pr{slucas}^\SStructure=1$ and $\pr{smith}^\SStructure=-1$,
\end{itemize}
and (iii) predicate symbols $\to^*,\to\:\mathrel{\in}\SPredicates_{\pr{WebPage}\:\pr{WebPage}}$ both interpreted as $\geq$ is a model of
$\ol{\cR}\cup\{\neg(\exists \mathtt{u:EventualUser}) ~\mathtt{wwv05(u)}\to^*\mathtt{submit(u)}\}$, thus proving the desired 
security property.
Note that this crucially depends on the type \pr{RegUser} of variable \pr{R} controling the `identity' of any user reaching the 
web page \pr{submit}.
If a rank \pr{submit : User -> SecureWebPage} is used instead of the current one but variable \pr{R} in the rule for \pr{submit} is of
type \pr{RegUser}, the property still holds.
However, if \pr{R} is of type \pr{User}, then no model is obtained.

\section{Related Work}\label{SecRelatedWork}

The so-called \emph{first-order theory of rewriting} (\emph{FOThR} in the following) 
uses a restricted first-order language (without constant or function symbols,
and with only two predicate symbols $\to$ and $\to^*$).
The predicate symbols are by definition \emph{interpreted} on an \emph{intended model} 
that, for a given TRS $\cR$,  gives meaning to 
$\to$ and $\to^*$ as the 
one-step and many-step rewrite relations $\to_\cR$ and
$\to^*_\cR$ for $\cR$ on \emph{ground terms}, respectively 
\cite{DauTis_TheTheoryOfGroundRewriteSystemsIsDecidable_LICS90}.
Note that this is just the least Herbrand model $\cH_\cR$ associated to the Horn theory $\ol{\cR}$ of $\cR$!
\emph{FOThR} is often used to express and verify properties of TRSs. 
For instance, confluence can be expressed as follows:
\begin{eqnarray}
(\forall x,y,z)~(x\to^*y\wedge x\to^*z\Rightarrow (\exists u) (y \to^*u\wedge z\to^*u))\label{FOThRsentenceForConfluence}
\end{eqnarray}
Given a TRS $\cR$ and a formula $\varphi$ in the language of \emph{FOThR}, $\cH_\cR\models\varphi$ 
(i.e., the satisfiability of $\varphi$ in $\cH_\cR$) actually \emph{means} that
the property expressed by $\varphi$ \emph{holds} for the TRS $\cR$.
For instance $\cH_\cR\models(\ref{FOThRsentenceForConfluence})$ means `$\cR$ is ground confluent'.
And $\neg(\cH_\cR\models(\ref{FOThRsentenceForConfluence}))$, which is equivalent to
$\cH_\cR\models\neg(\ref{FOThRsentenceForConfluence})$ means `$\cR$ is \emph{not} ground confluent'.
Decision algorithms for these properties exist for restricted classes of TRSs $\cR$ like 
left-linear right-ground TRSs, where variables are allowed in the left-hand side of the rules (without repeated occurrences of the
same variable) but disallowed in the right-hand side 
\cite{RapMid_AutomatingTheFirstOrderTheoryOfRewritingForLeftLinearRightGroundRewriteSystems_FSCD16}.
However, a simple fragment of \emph{FOThR} like the \emph{First-Order Theory of One-Step Rewriting}, where only a single
predicate symbol $\to$ representing one-step rewritings with $\cR$ is allowed, has been proved undecidable even for \emph{linear} TRSs
\cite{Treinen_FOTheoryRewritingUndecidable_TCS98}.

In contrast,
we use the full expressive power of first-order logic
to represent sophisticated rewrite theories where sorts, conditional rules and equations, membership predicates, etc., are allowed.
We do not impose any restriction on the class of rewrite systems we can deal with.
In contrast to \emph{FOThR}, where function symbols are not allowed in formulas, we can use \emph{arbitrary} sentences involving 
arbitrary terms.
Also in contrast to \emph{FOThR}, with a single allowed model $\cH_\cR$, 
we permit the \emph{arbitrary interpretation} of the underlying first-order logic language for proving properties.
As a consequence of this, though, we also need to impose restrictions to the shape of first-order sentences we can deal with meaningfully.
The application of this approach to well-known problems in rewriting leads to new methods which show their usefulness with regard
to existing methods.
In contrast to \emph{FOThR}, though, sentences like (\ref{FOThRsentenceForConfluence})
do not fit format (\ref{ExistentialClosureOfPositiveBooleanCombination}) considered in this paper
(but most sentences in Figure \ref{FigSomeFOpropertiesRewritingBasedSystems} cannot be expressed in 
\emph{FOThR} either, as they involve specific \emph{terms} with or without
variables).

Other approaches like the ITP tool, \emph{a theorem prover that can be used to prove properties of
membership equational specifications} \cite{ClaPalRie_IntroducingTheITPtoolATutorial_JUCS06}
work similarly:
the tool can be used to 
verify such properties \emph{with respect to ITP-models} which are actually special versions of the Herbrand model 
of the underlying theory. Then, one may have similar decidability problems as discussed for \emph{FOThR}.

\section{Conclusions and future work}\label{SecConclusions}

We have presented a semantic approach to prove properties of computational systems whose semantics can be given
as a Horn theory $\cS$. 
Provided that a program property can be expressed as a first-order sentence $\varphi$ which is 
the existential closure of a positive boolean combination of atoms, 
the \emph{satisfaction} of the negation $\neg\varphi$ of this sentence 
by an arbitrary model $\SStructure$ of $\cS$ implies that $\neg\varphi$ holds in the standard Herbrand model of $\cS$.
As usual, we can think of this fact as $\cS$ actually \emph{missing} the property expressed by $\varphi$.

We have explained how to apply this simple technique to deal with rewriting-based computational systems, in
particular with (possibly sorted) conditional rewrite systems.
We have considered a number of properties that have been investigated in the literature (infeasibility of conditional critical
pairs and rules, non-joinability of ground terms, non-loopingness, nonreachability, etc.).
Quite surprisingly, we could handle many specific examples coming from papers developing specific techniques to deal with these problems 
with our semantic approach (Corollary \ref{CoroSemanticApproach}).
In particular, we could deal with all the examples solved in \cite{SteMid_InfeasibleConditionalCriticalPairs_IWC15,%
SteSte_CertifyingConfluenceOfAlmostOrthogonalCTRSsViaExactTreeAutomataCompletion_FSCD16} 
(some of them reported in our examples above; note that these papers explore several \emph{alternative}
methods and, as reported by the authors, some of them \emph{fail} in specific examples which then require 
a different approach).
We also dealt with all Aoto's examples  in  \cite{Aoto_DisprovingConfluenceOfTermRewritingSystemsByInterpretationAndOrdering_FroCoS13} 
in combination with his \emph{usable rules} refinement (see also \cite{LucGut_ASemanticCriterionForProvingInfeasibitlityInConditionalRewriting_IWC17}).
Furthermore,  these examples were all handled by using our tool \AGES\ for the automatic generation of 
models of Order-Sorted First-Order Theories.

In the future, we plan to improve the ability of our methods to deal with more general properties.
In particular, a better use of sorts when modeling computational systems looks promising (as suggested in Section \ref{SecRunnningExampleSecureWeb}), 
in a similar way as \emph{type
introduction} improves the ability to prove properties of TRSs 
\cite{Zantema_TerminationOfTermRewIntAndTypeElim_JSC94}.

\medskip
\noindent
{\bf Acknowledgements.} I thank Mar\'{\i}a Alpuente and Jos\'e Meseguer for fruitful discussions about the topics in this paper.


{\small

\begin{thebibliography}{10}

\bibitem{Aoto_DisprovingConfluenceOfTermRewritingSystemsByInterpretationAndOrdering_FroCoS13}
T.\ Aoto.
\newblock Disproving Confluence of Term Rewriting Systems by Interpretation and Ordering. 
\newblock In
\emph{Proc.\ of
FroCoS'13}, 
LNCS 8152:311-326, 2013.

\bibitem{BaaNip_TermRewAllThat_1998}
F. Baader and T. Nipkow.
\newblock Term Rewriting and All That.
\newblock Cambridge University Press, 1998.

\bibitem{BerKlo_ConditionalRewriteRulesConfluenceAndTermination_JCSS86}
J.A. Bergstra and J.W. Klop.
\newblock Conditional Rewrite Rules: Confluence and Termination.
\newblock \emph{Journal of Computer and System Sciences} 32:323-362, 1986.

\bibitem{BruMes_SemFoundGRT_TCS06}
R.~Bruni and J.~Meseguer.
\newblock Semantic foundations for generalized rewrite theories.
\newblock {\em Theoretical Computer Science} 351(1):386-414, 2006.


\bibitem{ChaLee_SymbolicLogicAndMechanicalTheoremProving_1973}
C.L.\ Chang and R.C.\ Lee.
\newblock Symbolic Logic and Mechanical Theorem Proving.
\newblock Academic Press, 1973.

\bibitem{Clark_PredicateLogicAsAComputationalFormalism_TR79}
K.L.\ Clark.
\newblock Predicate Logic as a Computational Formalism.
\newblock PhD.\ Thesis, Research Monograph 79/59 TOC, Department of Computing, Imperial College of
Science, and Technology, University of London, December 1979.

\bibitem{ClavelEtAl_MaudeBook_2007}
M.~Clavel, F.~Dur\'an, S.~Eker, P.~Lincoln, N.~Mart\'{\i}-Oliet, J.~Meseguer,
  and C.~Talcott.
\newblock All About Maude -- A High-Performance Logical Framework.
\newblock LNCS 4350, Springer-Verlag, 2007.

\bibitem{ClaPalRie_IntroducingTheITPtoolATutorial_JUCS06}
M.~Clavel, M.\ Palomino, and A.\ Riesco.
\newblock Introducing the ITP Tool: a Tutorial.
\newblock \emph{Journal of Universal Computer Science} 12(11):1618-1650, 2006.


\bibitem{DauTis_TheTheoryOfGroundRewriteSystemsIsDecidable_LICS90}
M.\ Dauchet and S.\ Tison.
\newblock The Theory of Ground Rewrite Systems is Decidable.
\newblock In \emph{Proc.\ of 
LICS '90}, 
pages 242-248, IEEE Press, 1990

\bibitem{Dershowitz_TerminationOfRewriting_JSC87}
N. Dershowitz.
\newblock Termination of rewriting.
\newblock {\em Journal of Symbolic Computation}, 3:69-115, 1987.

\bibitem{DerOka_RationaleForConditionalEqProgramming_TCS90}
N. Dershowitz and M. Okada.
\newblock A rationale for conditional equational programming.
\newblock {\em Theoretical Computer Science} 75:111-138, 1990.

\bibitem{DuranEtAl_ProvOpTermMEqProg_HOSC08}
F.~Dur\'{a}n, S.~Lucas, C.~March\'{e}, J.~Meseguer, and X.~Urbain.
\newblock {Proving Operational Termination of Membership Equational Programs}.
\newblock {\em Higher-Order and Symbolic Computation}, 21(1-2):59--88, 2008.


\bibitem{EmdKow_TheSemanticsOfPredicateLogicAsAProgrammingLanguage_JACM76}
M.H.\ van Emden and R.A.\ Kowalski.
\newblock The semantics of Predicate Logic as a Programming Language.
\newblock \emph{Journal of the ACM} 23(4):733-742, 1976.

\bibitem{GaiHilLocSpi_TheNewWaldmeisterLoopAtWork_CADE03}
J.-M.\ Gaillourdet, T.\ Hillenbrand, B.\ L\"ochner, and H.\ Spies.
\newblock The New WALDMEISTER Loop at Work.
\newblock In
\emph{Proc. of
CADE'03},
LNCS,  2741:317-321, 2003.
    
\bibitem{GieArt_VerificationErlangProcessesDPs_AAECC01}
J.~Giesl and T. Arts.
\newblock Verification of Erlang Processes by Dependency Pairs.
\newblock \emph{Applicable Algebra in Engineering, Communication and Computing}
12:39-72, 2001.
%
\bibitem{GogMes_ModelsAndEqualityForLogicalProgramming_TAPSOFT87}
J. Goguen and J. Meseguer.
\newblock Models and Equality for Logical Programming.
\newblock In 
\emph{Proc.\ of
TAPSOFT'87}, 
LNCS 250:1-22, Springer-Verlag, 1987.

\bibitem{GutLucRei_AToolForTheAutomaticGenerationOfLogicalModelsForOrderSortedFirstOrderTheories_PROLE16}
R. Guti\'errez, S. Lucas, and P. Reinoso.
\newblock A tool for the automatic generation of logical models of order-sorted first-order theories.
\newblock In 
\emph{Proc. of 
PROLE'16}, pages 215-230, 2016. \url{http://hdl.handle.net/11705/PROLE/2016/018}.
Tool available at \url{http://zenon.dsic.upv.es/ages/}.

\bibitem{Hodges_AShorterModelTheory_1997}
W. Hodges.
\newblock A shorter model theory.
\newblock Cambridge University Press, 1997.

\bibitem{Lucas_RewritingBasedNavigationOfWebSites_WWV05}
S. Lucas.
\newblock Rewriting-based navigation of web sites.
\newblock In {\em Proc. of 
WWV'05}, ENTCS 157:79-85,  2006.

\bibitem{LucGut_ASemanticCriterionForProvingInfeasibitlityInConditionalRewriting_IWC17}
S.\ Lucas and R.\ Guti\'errez.
\newblock A Semantic Criterion for Proving Infeasibitlity In Conditional Rewriting.
\newblock In 
\emph{Proc of
IWC'17},
pages 15-20, 2017.

\bibitem{LucGut_AutomaticSynthesisOfLogicalModelsForOrderSortedFirstOrderTheories_JAR17}
S.~Lucas and R.\ Guti\'errez.
\newblock Automatic Synthesis of Logical Models for Order-Sorted First-Order Theories.
\newblock {\em Journal of Automated Reasoning}, DOI 10.1007/s10817-017-9419-3, 2017.
%
\bibitem{LucMarMes_OpTermCTRSs_IPL05}
S.~Lucas, C.~March\'e, and J.~Meseguer.
\newblock Operational termination of conditional term rewriting systems.
\newblock {\em Information Processing Letters} 95:446--453, 2005.
%
\bibitem{LucMes_ModelsForLogicsAndConditionalConstraintsInAutomatedProofsOfTermination_AISC14}
S.~Lucas and J.~Meseguer.
\newblock Models for Logics and Conditional Constraints in Automated Proofs of Termination.
\newblock In 
{\em Proc. of  
AISC'14}, 
LNAI 8884:9-20, 2014.

\bibitem{LucMes_DependencyPairsForProvingTerminationPropertiesOfCTRSs_JLAMP17}
S.~Lucas and J.~Meseguer.
\newblock Dependency pairs for proving termination properties of conditional
term rewriting systems.
\newblock \emph{Journal of Logical and Algebraic Methods in Programming}, 86:236-268, 2017.

\bibitem{LucMes_NormalFormsAndNormalTheoriesunconditionalRewriting_JLAMP16}
S.~Lucas and J.~Meseguer.
Normal forms and normal theories in conditional rewriting.
\newblock \emph{Journal of Logical and Algebraic Methods in Programming}, 85(1):67-97, 2016.

\bibitem{LucMesGut_ExtendingThe2DDPFrameworkForCTRSs_LOPSTR14}
S.~Lucas, J.~Meseguer, and R. Guti\'errez.
\newblock Extending the 2D DP Framework for Conditional Term Rewriting Systems.
\newblock In 
\emph{Selected papers of 
LOPSTR'14},  LNCS 8981:113-130, 2015.

\bibitem{Mendelson_IntroductionToMathematicalLogicFourtEd_1997}
E.\ Mendelson.
\newblock Introduction to Mathematical Logic. Fourth edition.
\newblock Chapman \& Hall, 1997.

\bibitem{Meseguer_MembershipAlgebraAsALogicalFrameworkForEquationalSpecification_WADT97}
J.~Meseguer.
\newblock Membership algebra as a logical framework for equational
  specification.
\newblock In 
{\em Proc. of 
WADT'97}, LNCS 1376:18--61, Springer-Verlag, 
1998.

\bibitem{Meseguer_20YearsRewLogic_JLAP12}
J.~Meseguer.
\newblock Twenty years of rewriting logic.
\newblock {\em Journal of Logic and Algebraic Programming} 81:721-781, 2012.


\bibitem{Ohlebusch_AdvTopicsTermRew_2002}
E.~Ohlebusch.
\newblock {\em Advanced Topics in Term Rewriting}.
\newblock {Sprin\-ger-Verlag}, Apr. 2002.

\bibitem{Prawitz_NaturalDeductionAProofTheoreticalStudy_2006}
D.\ Prawitz.
\newblock Natural Deduction. A proof-theoretical study. Dover, 2006.


\bibitem{RapMid_AutomatingTheFirstOrderTheoryOfRewritingForLeftLinearRightGroundRewriteSystems_FSCD16}
F. Rapp and A.\ Middeldorp.
\newblock Automating the First-Order Theory of Rewriting for Left-Linear Right-Ground Rewrite Systems.
\newblock In 
\emph{Proc.\ of 
FSCD'16}, LIPIcs 52, Article No. 36; pp. 36:1?36:12, 2016.

\bibitem{Smullyan_TheoryOfFormalSystems_1961}
R.M.\ Smullyan.
\newblock Theory of Formal Systems.
\newblock Princeton University Press, 1961.

\bibitem{SteMid_ConditionalConfluenceSystemDescription_RTATLCA14}
T.\ Sternagel and A.\ Middeldorp.
\newblock Conditional Confluence (System Description).
\newblock In
\emph{Proc. of
RTA-TLCA'14},
LNCS 8560:456-465, 2014.

\bibitem{SteMid_InfeasibleConditionalCriticalPairs_IWC15}
T.\ Sternagel and A.\ Middeldorp.
\newblock Infefasible Conditional Critical Pairs.
\newblock In
\emph{Proc. of
IWC'15},
pages 13--18, 2014.

\bibitem{SteSte_CertifyingConfluenceOfAlmostOrthogonalCTRSsViaExactTreeAutomataCompletion_FSCD16}
C.\ Sternagel and T.\ Sternagel.
\newblock Certifying Confluence Of Almost Orthogonal CTRSs Via Exact Tree Automata Completion.
\newblock In 
\emph{Proc.\ of 
FSCD'16}, LIPIcs 52, Article No. 85; pp. 85:1--85:16, 2016.

\bibitem{Treinen_FOTheoryRewritingUndecidable_TCS98}
R.\ Treinen.
\newblock The first-order theory of linear one-step rewriting is undecidable.
\newblock \emph{Theoretical Computer Science} 208:179-190, 1998.

\bibitem{Wang_LogicOfManySortedTheories_JSL52}
H. Wang.
\newblock Logic of many-sorted theories.
\newblock \emph{Journal of Symbolic Logic} 17(2):105-116, 1952.

\bibitem{Zantema_TerminationOfTermRewIntAndTypeElim_JSC94}
H.~Zantema.
\newblock Termination of term rewriting: interpretation and type elimination.
\newblock {\em Journal of Symbolic Computation}, 17:23-50, 1994.


\end{thebibliography}}
\end{document}